\documentclass[12pt]{article}

\usepackage{xr}
\usepackage{graphicx}
\usepackage[dvipsnames]{xcolor}
\usepackage{soul}
\usepackage[normalem]{ulem}
\usepackage[letterpaper,margin=1in]{geometry}

\usepackage{siunitx}

\usepackage{amsmath} 
\usepackage{soul}
\usepackage{amsfonts}
\usepackage{amssymb}
\usepackage{physics}
\usepackage{color, soul}
\usepackage{graphicx}
\usepackage[caption=false]{subfig}
\usepackage{array}
\usepackage{booktabs}
\usepackage{xcolor}
\usepackage{times}
\usepackage{float}
\UseRawInputEncoding
\usepackage[numbers,sort&compress]{natbib}

\usepackage{newtxtext,newtxmath}

\newif\ifhidecomments
\hidecommentstrue
\ifhidecomments
\newcommand{\JP}[1]{#1}

\newcommand{\MS}[1]{#1}

\else
\newcommand{\JP}[1]{\textcolor{blue}{#1}}

\newcommand{\MS}[1]{\textcolor{Plum}{#1}}
\definecolor{mickaelcolor}{rgb}{0.35,0.53,0.9}

\fi

\usepackage{gensymb}
\newcommand{\C}{\degree\textrm{C}}
\renewcommand{\parallel}{\mathbin{\!/\mkern-5mu/\!}}
\renewcommand{\epsilon}{\varepsilon}
\newenvironment{sciabstract}{%
\begin{quote} \bf}
{\end{quote}}

\usepackage{xspace}
\newcommand{\n}{\textbf{n}\xspace}

\newcounter{lastnote}

\title{\selectfont \textbf{Scale-coupling from kirigami cuts controls emergent mechanics in liquid crystal elastomers}}
\author{
	M. Strugaru$^{1}$,
	M. Ly$^{1}$,
	Q. Martinet$^{1}$,
   B. Bickel$^{2}$, 
    J. Palacci$^{1\ast}$\and
	\small$^{1}$Institute of Science and Technology Austria, Klosterneuburg, Austria \\
	\small$^{2}$ETH Z{\"u}rich\and
\small$^\ast$ Corresponding author. Email: jeremie.palacci@ista.ac.at\and
}

\begin{document}
%\baselineskip16pt

% Make the title.

\maketitle

\begin{sciabstract}
%\JP{
Conventional materials derive properties from microscopic composition and arrangement. On the contrary, mechanical metamaterials are engineered materials whose unconventional properties are governed by their mesoscopic structure, not by what they are made of. To date, the two paradigms have remained decoupled, and how macroscopic geometric alterations can orchestrate microscopic degrees of freedom to program mechanics is a central challenge. Here, we demonstrate that cuts in anisotropic, responsive solids provide such a link between scales. 
Using sheets of liquid crystal elastomer (LCE), patterned with cuts -- or LCE kirigami,  we reveal that the programming of strain by cuts harnesses the molecular anisotropy of LCEs to control emergent mechanics. Similarly, the interplay of cut patterns with the molecular phase transition of LCEs gives rise to soft robotic functions such as supersoft grippers with remote actuation or multifarious architectures that reversibly morph with changes in temperature patterns -- all features inaccessible to conventional kirigami or LCE sheets alone. Through LCE kirigami, we showcase a new class of multiscale metamaterials in which geometry governs access to microscopic  degrees of freedom to program macroscopic function. 
\end{sciabstract}

\section*{Introduction}
\begin{figure}[htbp]
\centering
\includegraphics[width=\linewidth]{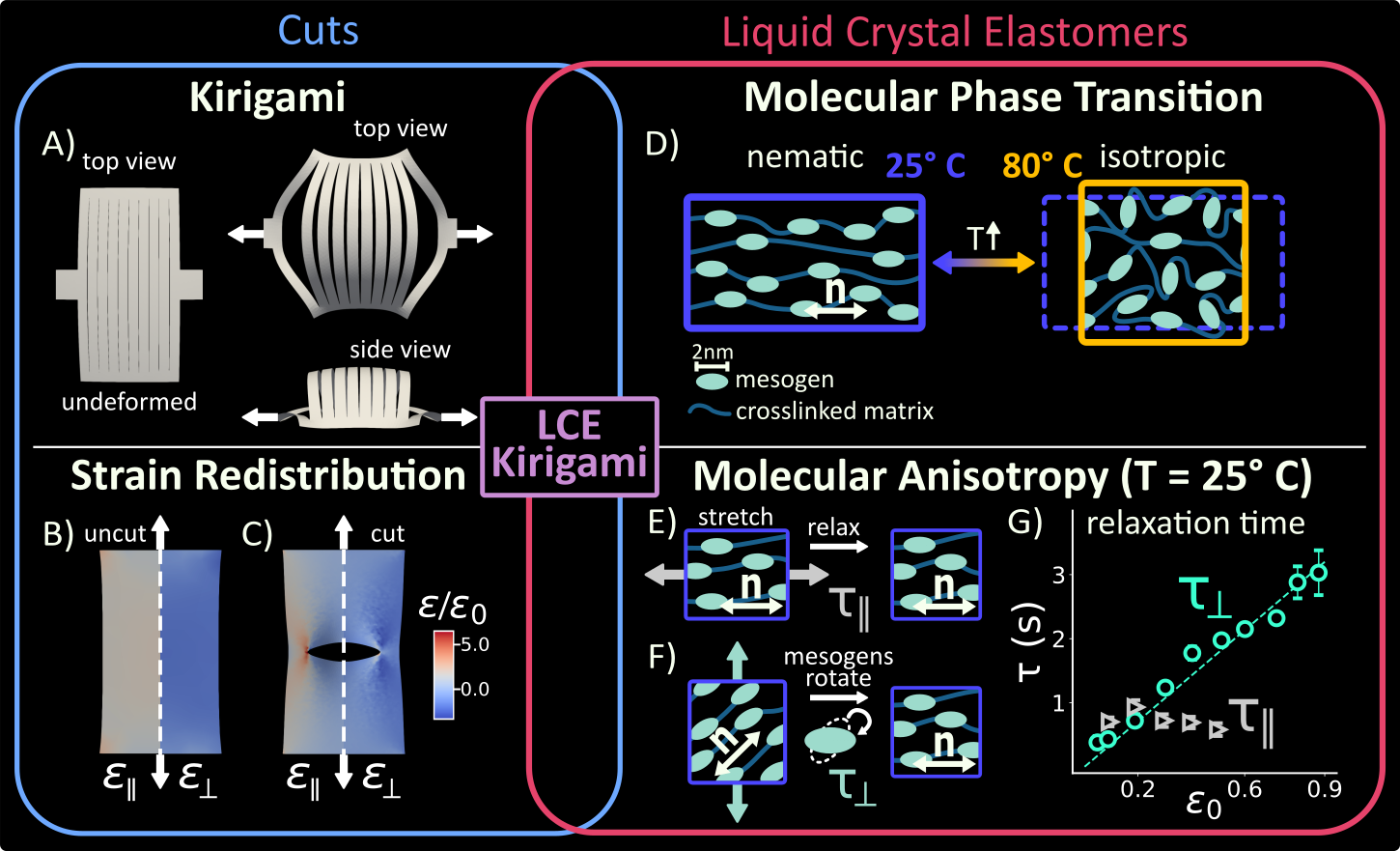}
\caption{ {\bf LCE kirigami are multiscale metamaterials.} The interplay of mesoscale alterations (cuts, A-C) with the molecular structure of Liquid Crystal Elastomers - LCEs (D-F), dubbed LCE kirigami,  crosses scales to form a multiscale metamaterial.  { \bf  A-C)  Properties of cuts.} A) Patterns of cuts or kirigami are a versatile basis for metamaterials, here showing a 3D deformation from a 2D sheet. B-C) Simulated parallel, $\varepsilon_{\parallel}$, and perpendicular, $\varepsilon_{\perp}$, strains for thin sheets under uniaxial global strain $\epsilon_0$. B) Uncut sheets present uniform strains $\varepsilon_{\parallel}$ and  $\varepsilon_{\perp}$in both directions, while C) the presence of cut redirects and amplifies strain,  concentrating  $\varepsilon_{\perp}$ near the cut tip. Colorbar indicates the relative strain $\epsilon/\epsilon_0$. {\bf D-G) Molecular properties of Liquid Crystal Elastomers (LCEs).} Nematic director \n indicated by the white double arrow for all LCEs. D) LCE sheet: a crosslinked elastomer matrix with embedded liquid crystal mesogens, with a uniform nematic director {\bf n} at room temperature. The molecular nematic-isotropic transition of the mesogens results in $30\%$ contraction along {\bf n} and $20\%$ expansion orthogonally from $25\C$ to $80\C$. 
%{\bf E-G) The nematic order at room temperature results in mechanical anisotropy.} 
E) LCE sheets stretched along {\bf n} retract elastically with the elastomer matrix. F) LCE sheets stretched perpendicular to {\bf n} (light blue arrows) result in the rotation of the  mesogens, linking molecular order to macroscopic strain. After release, they rotate back. G) Measurements of the relaxation time $\tau$ following stretch along {\bf n}, $\tau_{//}$ (gray triangles), and perpendicular to {\bf n}, $\tau_{\perp}$ (light blue circles). The interplay between macroscopic strain and molecular anisotropy results in $\tau_{\perp}\propto \epsilon_0$, while $\tau_{\parallel}\lessapprox 1s$ as for an isotropic elastomer. 
}
\label{fig:figure1}
\end{figure}

In material science, mechanical properties are governed by  materials' constituents, stressing the link between microscopic structure and macroscopic behavior. This contrasts with the central paradigm in mechanical metamaterials: that properties emerge from the mesoscopic geometric structure, not from material composition.  
%\JP{
The latter led to unconventional mechanics in numerous fashions such as auxetic behavior \cite{10.1115/1.4031456, 10.1038/s41467-019-12757-7}, functional \cite{10.48550/arxiv.2506.18197, 10.1002/advs.202001271, 10.1103/physrevx.8.021075, 10.1126/science.aaz0135, 10.1103/physrevlett.115.118302, 10.1038/s41467-025-61809-8}, or shape-morphing \cite{10.1038/s41586-023-06353-5, SOAmeta25, SOAmeta22, SOAmeta17} materials.
%} 
%Architectures of slender objects present auxetic behavior (dilation under extension) following the buckling of beams \cite{10.1115/1.4031456, 10.1038/s41467-019-12757-7}. Commonplace trades such as weaving \cite{10.48550/arxiv.2506.18197, 10.1002/advs.202001271}, knitting \cite{10.1103/physrevx.8.021075}, or knotting \cite{10.1126/science.aaz0135, 10.1103/physrevlett.115.118302} employ non-linear  mechanics and topology to create functional materials. The geometry, tension, and friction of threads of rigid beads program the mechanics of their assembly \cite{10.1038/s41467-025-61809-8}. Meanwhile,traditional handicrafts such as origami, kirigami cuts, or pleats \cite{10.1038/s41586-023-06353-5} are reinterpreted as metamaterials from 2D templates. 
%\JP{To date, metamaterials and molecular materials have remained largely decoupled design paradigms.
%\JP{
Aiming to thread mesoscopic alterations with molecular features of responsive materials [Fig.~\ref{fig:figure1}], we investigate the mechanics of 2D sheets of Liquid Crystal Elastomers (LCEs) with cut patterns, or {\it LCE kirigami}. We demonstrate that macroscopic cuts selectively activate molecular degrees of freedom, coupling scales over 7 orders of magnitude, from $nm$ to $cm$, to program emergent mechanics and make LCE kirigami more than the sum of their parts.
%}
\\

First, cuts provide a bridge between the mesoscopic and molecular arenas by orienting and concentrating stress down to the microscopic scale [Fig.~\ref{fig:figure1}B-C] \cite{stress_plate, tear_thinsheet}. In addition, patterns of cuts - kirigami -  convert simple input, e.g. uniaxial stretch, into complex spatiotemporal strains and constitute a versatile basis for metamaterials: controlling mechanical properties  \cite{microstruct, diasp}, shape-morphing structures \cite{kiricurvature, kirigami3Ddefs, kirigrip} [Fig.~\ref{fig:figure1}A], or dynamical behavior  \cite{liangrec, bistableviscoel}.
Second, Liquid Crystal Elastomers (LCEs) are cross-linked polymer networks with embedded liquid crystal mesogens [Fig.~\ref{fig:figure1}D], a hybrid architecture that confers them with remarkable and controllable mechanical properties \cite{lcebook,10.1021/acs.macromol.4c01997}. The molecular ordering of the mesogens is characterized by a local nematic director ${\bf n}$: mesogens are aligned during synthesis with their orientation fixed by a curing step \cite{lcefab, lce4Dprint}.
Importantly, LCEs present a nematic-isotropic transition controlled by temperature, meaning that the nematic order, existing at room temperature, vanishes at $T\sim 80\C$. Because mesogens are tethered to the elastomeric network, this molecular phase transition is accompanied by a macroscopic contraction along the director and expansion in the orthogonal direction,  making  LCEs a material of choice for artificial muscles in soft robotics \cite{10.1021/acs.macromol.4c01997, elephant_trunk, liu2021lcegrids}. For this work, we consider 2D LCE sheets 
uniaxially stretched during synthesis [SI Sec.~S2],  setting a uniform director {\bf n} at room temperature. Increasing the temperature to $80\C$ results in $\sim 30\%$ contraction along {\bf n} and $\sim 20\%$ expansion in the orthogonal direction [Fig.~\ref{fig:figure1}D, SI Sec.~S3.1].
The molecular order further implies an anisotropic mechanical response of LCEs at room temperature [SI Sec.~S3.2]. When the LCE is extended uniaxially along the director,  mesogens are pulled with the stretch of the polymer matrix and LCEs respond elastically. In contrast, strains perpendicular to the director [Fig.~\ref{fig:figure1}E] require the reorientation and rotation of the mesogens in the polymer matrix [Fig.~\ref{fig:figure1}F], a complex microscopic rearrangement, which requires time. This leads to non-linear mechanics, known as soft elasticity \cite{lcebook, lianglc}, and results in strain-rate dependent behavior \cite{terentjev_dynandrelax}.

To investigate how strain controls the rotation of the mesogens, we study the elementary setting, whereby anisotropic (room temperature) LCE sheets are stretched uniaxially with increasing initial strain $\epsilon_0$ [Fig. 1G]. After release, the LCE sheet retracts with a relaxation time $\tau$, defined as the time to recover $80\%$ of the initial strain [Fig.~S1, Fig. S2]. LCE sheets stretched along the director retract rapidly with near constant time $ \tau_{\parallel}\lessapprox 1s$, reflecting the faster elastic response of the polymer matrix [Fig. 1G, gray triangles]. When stretched perpendicular to the director {\bf n}, the relaxation process slows down considerably, scaling with initial strain, $\tau_\perp\propto \epsilon_0$ [Fig.~\ref{fig:figure1}G, blue circles].
Higher strains increase the rotation of the mesogens, which rotate back during release, leading to a slower retraction, as observed in the non-linear rheology of LCEs \cite{terentjev_dynandrelax, coupling_viscoel_rotation}. This difference vanishes at $80\C$, where the LCE sheets behave isotropically and retract rapidly in both directions, acting effectively like elastomer sheets [Fig.~S3-inset].
We now unveil emergent mechanics arising from the scale coupling between cuts [Fig.~\ref{fig:figure1}A-C] and the molecularly-driven properties of LCEs [Fig.~\ref{fig:figure1}D-G]. 

\section*{Results}
\subsection*{Cut patterns control mesogen rotation and dynamics of LCE kirigami}

\subsubsection*{Cut patterns control heterogeneous strain and dynamics of LCE kirigami}

\begin{figure}[tbhp]
\centering
\includegraphics[width=0.95\linewidth]{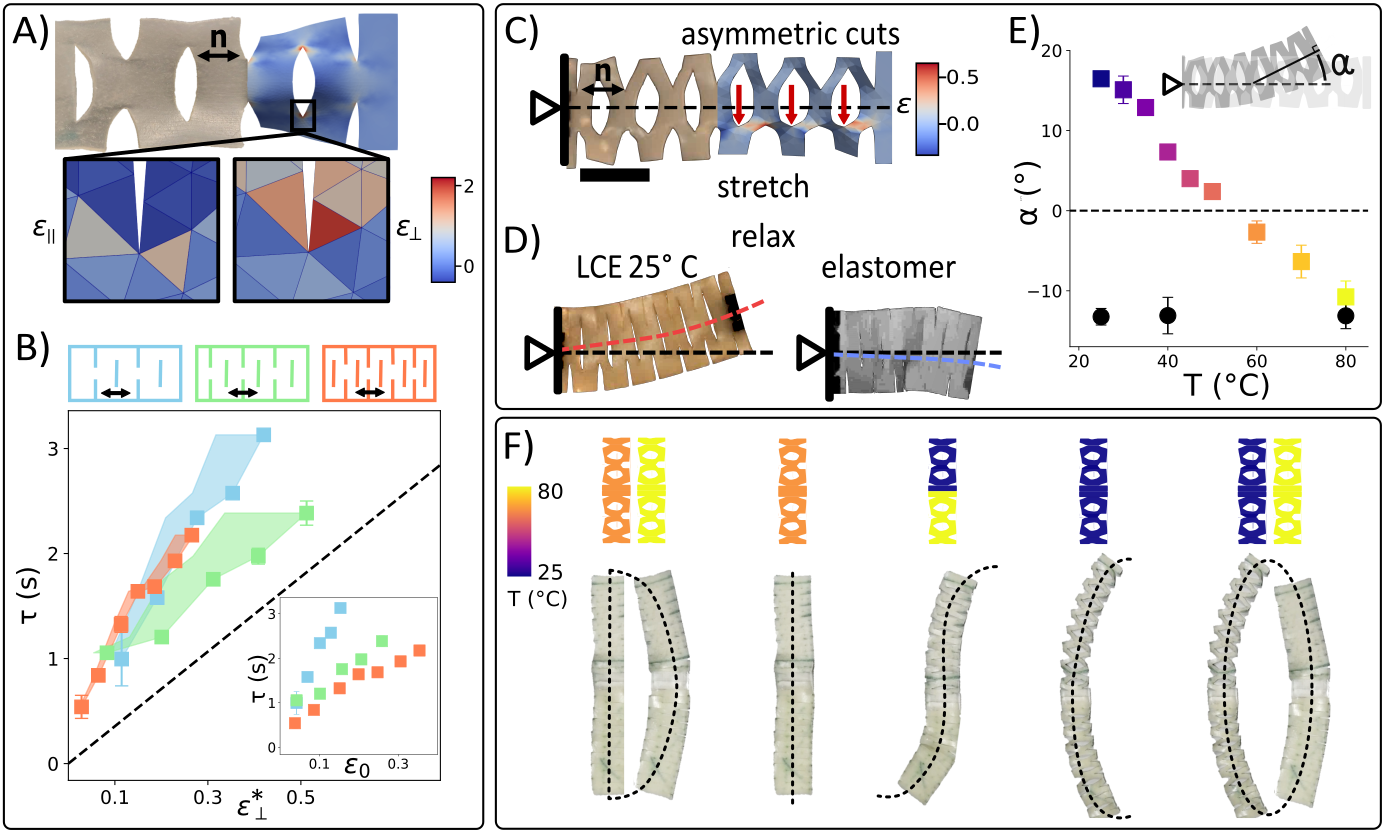}
\caption{{\bf Cut patterns control the dynamics of relaxation of  LCE kirigami.}
%The molecular order of LCEs controls relaxation dynamics and"pull-to-shape" programmability.
{\bf A-B) Dynamics of relaxation of LCE kirigami after uniaxial stretch along {\bf n} at T$=25\C$.} 
A) LCE kirigami under uniaxial \JP{strain $\epsilon_0\sim 0.25$} (left, experiment) and local strain (right, simulations). Cuts reorient and accumulate strain near the cut tip. Simulated   parallel, $\varepsilon_{\parallel}$ [left zoom-in], and perpendicular, $\varepsilon_{\perp}$ [right zoom-in], strain. Colorbar indicates the simulated strain with $\epsilon_0\sim 0.25$. (B-inset) Relaxation time $\tau$ of LCE kirigami (patterns on first row) to recover $80\%$ of initial strain $\epsilon_0$,  showing $\tau\propto \epsilon_0$ and decreasing with increased density of cuts. 
B) Collapse of experimental measurements (inset) with   $\tau(\epsilon^*_{\perp})$, with $\epsilon^*_{\perp}$  maximal orthogonal strain from simulations [see main text].  Dashed line: $\tau_{\perp}(\epsilon_0)$ for uncut LCE sheets stretched $\perp {\bf n}$.  Shaded areas indicate error from numerical estimates of strain. %The collapse establishes that cuts control heterogeneous strain and select the molecular relaxation and dynamics of LCE kirigami [see main text].  
C) {\bf Programmable recoil of LCE kirigami with asymmetric cuts.} 
Kirigami with an asymmetric cut pattern is stretched (left) and simulated $\epsilon_{\perp}$ strain (right) with higher strain on the short side of the cut (red arrows). After release and recoil, the kirigami presents a bend $\alpha$ from the asymmetric strain and strain-dependent  relaxation. D) LCE kirigami at $25\C$ bend with $\alpha\sim 16^\circ$ opposite to conventional elastomer kirigami. E) Bending angle $\alpha$  for different temperatures, showing reversal of chirality for LCE kirigami (squares) at $T\sim 60^\C$ and independent of temperature for an elastomer kirigami (black circles). 
F) {\bf "Pull-to-Shape" programmability:}  LCE kirigami with identical asymmetric cuts are pulled, and released under different temperature patterns (first row), to display the letters of the word "DISCO" [see main text].
}
\label{fig:figure2}
\end{figure}

Following the dynamics of relaxation of uncut anisotropic (room temperature) LCE sheets [Fig. 1G], we turn to the effect of patterns of cuts [Fig. S4] perpendicular to the director ${\bf n}$ in a similar setting. We measure the relaxation time $\tau$ of such LCE kirigami after initial strain $\epsilon_0$  along the director \n [Fig.~2A]. 
%\JP{
Remarkably, $\tau$ increases with the initial strain $\epsilon_0$ [Fig.~\ref{fig:figure2}B-inset], a strain-dependent behavior similar to uncut LCEs stretched in the orthogonal direction [Fig.~1G, blue circles], and contrasting with uncut LCEs similarly stretched along {\bf n} [Fig.~1G, grey triangles]. This behavior, absent from standard elastomers [Fig. S5], is indicative of the role of cuts in redistributing global strain in anisotropic materials. In addition,  $\tau$ decreases with the number of cuts for a given initial stretch $\epsilon_0$  [Fig.~\ref{fig:figure2}B-inset]. We understand this behavior as a result of the interplay between the rotation of the mesogens 
%}
[Fig.~\ref{fig:figure1}F] and the strain set by cut patterns [Fig.~\ref{fig:figure2}A]. To test our hypothesis, we compare the relaxation times $\tau$ measured for the LCE kirigami with the ones previously measured for uncut LCE {\it under uniform strain}. This requires the determination of the local strain for different cut patterns, considering the molecular order of LCE sheets at room temperature. Cuts accumulate strain near the tips down to small scales, making direct measurements with sufficient resolution inaccessible. Instead, we adapt a Koiter shell simulation framework, accounting for the mechanical anisotropy of LCEs with an orthotropic material model \cite{Ciarlet05}, and compute the local parallel, $\epsilon_{\parallel}$, and orthogonal, $\epsilon_{\perp}$, strains for an LCE kirigami globally stretched along the director, 
%\JP{
while neglecting 3D bending %}
[Fig.~\ref{fig:figure2}A][SI Sec. S4]. 
%\JP{
Simulations confirm the emergence of strain orthogonal to the uniaxial stretch, leading to the rotation of the mesogens and controlling the dynamics of relaxation, providing an intuitive mechanism for the observed phenomenology.
%}
%We hypothesize that the strain-dependent behavior [Fig. 1G] and rotation can be approximated by considering only the slowest mode, {\it i.e.} maximal orthogonal strain in the stretched LCE kirigami.} 
In principle, the tip of each cut is a singularity, where stress and strain concentrate. Experimentally, this accumulation of stress results in the plastic deformation of the tip into circular shapes with radius $\sim 1$\,mm observed over multiple samples [SI Sec. S4.5]. We emulate the truncation from plasticity in simulations by averaging the local strain over regions of comparable area, extracting an average strain $\epsilon^*$ [SI Sec. S4.4]. 
Following, we show that the relaxation time of the LCE kirigami sheets under uniaxial strain $\epsilon_0$ can be predicted by identifying the highest orthogonal strain $\epsilon_\perp^*$ which controls the slowest relaxation mode of the sheet. As demonstrated in [Fig.~\ref{fig:figure2}B, Fig. S2], we see that that $\tau(\epsilon^*_{\perp})$ collapses our data for all cut patterns. Moreover, the time $\tau(\epsilon^*_{\perp})$ matches the relaxation time of an \emph{uncut} LCE sheet stretched orthogonally to the director with \emph{global} strain equal to the value of $\epsilon^*_{\perp}$, [Fig.~\ref{fig:figure2}B - dashed line], establishing quantitatively that mesoscale geometry selects the dominant molecular relaxation mode.
%\JP{
%Following, we predict the relaxation time of the LCE kirigami for increasing strain $\epsilon_0$, by making the approximation that it is selected by the slowest relaxation mode, {\it i.e.} the region of highest orthogonal strain $\epsilon_\perp^*$ and show that $\tau(\epsilon^*_{\perp})$ collapses our data for all cut patterns [Fig.~\ref{fig:figure2}B, Fig. S2]. It is set by the relaxation time of uncut LCE sheets stretched orthogonally to the director with global strain $\epsilon_0$  [Fig.~\ref{fig:figure2}B - dashed line], establishing quantitatively that mesoscale geometry selects the dominant molecular relaxation mode.
%} 

%We note that we neglected three dimensional bending and only considered the highest strain (slowest mode) in the orthogonal direction, two important simplifications in our model, which could be included in refinements to further improve the collapse of the experimental data. }

\subsubsection*{Pull-to-shape: programmable retraction of LCE kirigami.}
%\JP{
It results that heterogeneous strain creates spatially heterogeneous relaxation times, which we employ to produce controllable recoil curvatures and "pull to shape" programmability of LCE kirigami -- a feature inaccessible to conventional kirigami.
%}
%We now show that this interplay between cuts and strain-dependent relaxation time of LCEs programs the bending relaxation and shape of LCE kirigamis after stretch - "pull to shape", a feature inaccessible to conventional kirigami.
%\JP{
To this end, we consider an asymmetric pattern of cuts, the central cut shifted from the centerline, leading to asymmetric strain on each side of the centerline [Fig.~\ref{fig:figure2}C]. We contrast an LCE kirigami with an elastomer sheet of comparable elasticity with the same cut pattern [Fig.~\ref{fig:figure2}C-F].
%} 
% The central cut is shifted from the centerline, which leads to asymmetric strain under uniaxial strain - the short side of the cut is under more strain than the longer side, as computed in [Fig.~\ref{fig:figure2}C]. 
Following release, the LCE kirigami bends significantly towards the longer side of the cut, with $\alpha\sim 16^\circ$ [Fig.~\ref{fig:figure2}D], a behavior opposite to the same kirigami pattern in an elastomer sheet, where $\alpha<
  0$. Increasing the temperature reduces the bend for the LCE kirigami before the chirality reverses at $T\sim 60\C$, while $\alpha$ remains near constant for the elastomer kirigami [Fig. 2E]. These results illustrate that cuts reorient and distribute stress and strain locally,  amplifying the molecular anisotropy of the LCE sheet towards macroscale mechanics. We used this mechanism to program the retraction of LCE kirigami using temperature patterns. We call this effect "pull-to-shape", as illustrated in [Fig.~\ref{fig:figure2}F], 
  %\JP{
  where we write "DISCO" from asymmetric and identical LCE kirigami, reversibly programmed into different letters by temperature patterns [movie S1].
  %} 
  Taken together, our results show that mesoscopic cuts bridge scale and link macroscopic patterns with molecular properties of LCEs. We now reveal a new facet of the scale-coupling of LCE kirigami by turning the molecular nematic-isotropic transition of LCEs into soft robotics functions.

\subsection*{Molecular phase transition controls the remote actuation of LCE kirigami }

LCE sheets contract in the direction of the director \n when the temperature is increased above room temperature [Fig.~\ref{fig:figure1}D]\cite{liu2021lcegrids}, a result of the nematic-isotropic phase transition of the liquid crystal mesogens, reflected in optical measurements of birefringence [SI Sec.~S3.3]. Untethered LCE sheets with macroscopic cuts present the same contraction/expansion as uncut LCE sheets [SI Sec.~S3.4], confirming \JP{decoupling of scale between mesoscopic cuts and molecular order {\it in the absence of strain}}. This leads to a remarkable phenomenon when the opposite ends of an LCE kirigami are tethered in the direction normal to \n. When heated, the LCE sheet contracts and the LCE kirigami deploys [Fig.~\ref{fig:figure3}A] like an elastomer kirigami under uniaxial load [Fig.~\ref{fig:figure3}B]. 

\subsubsection*{Thermal Actuation of Kirigami}
\begin{figure}[htbp]
\centering
\includegraphics[width=\linewidth]{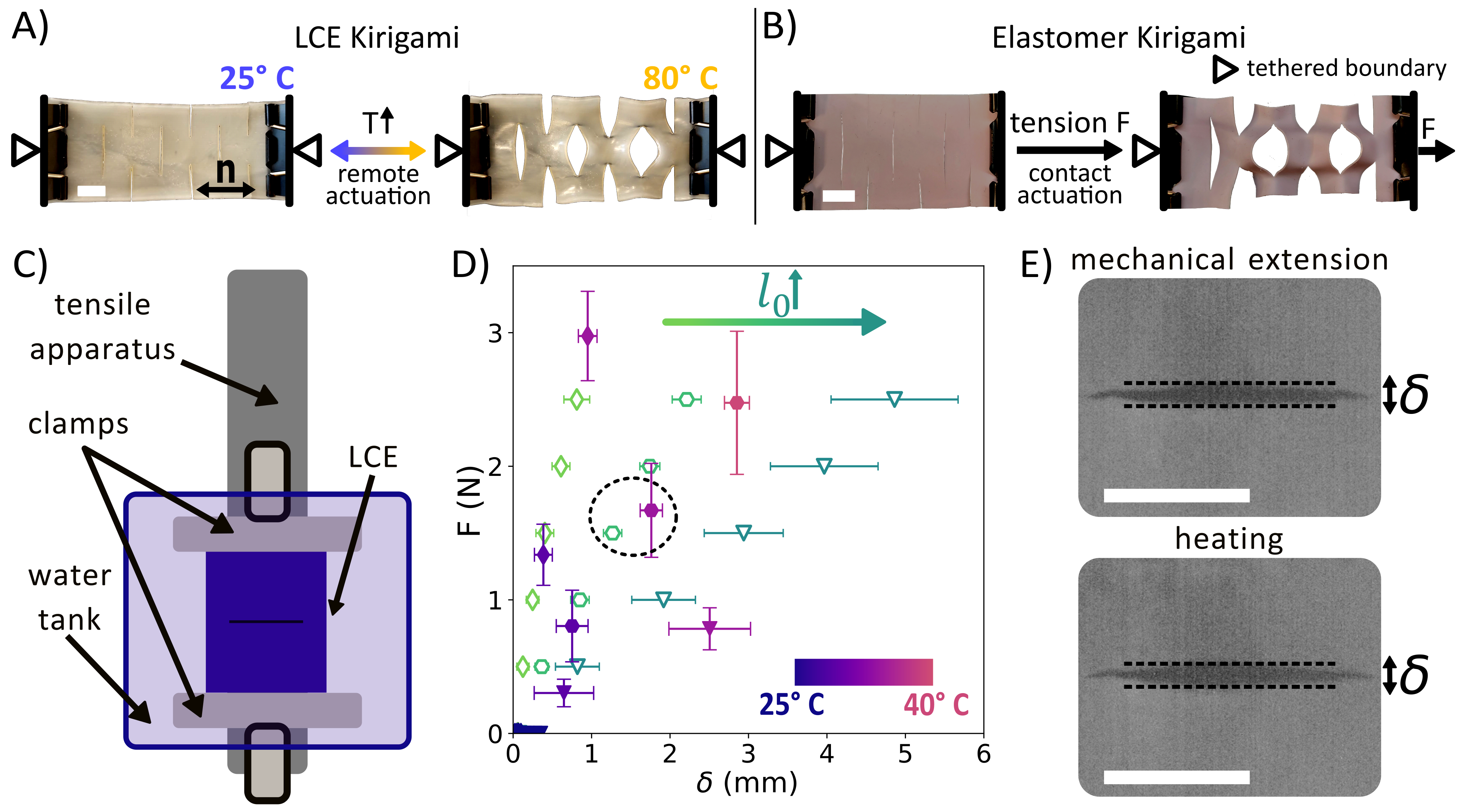}
\caption{
{\bf Actuation of LCE kirigami by pulling or heating-up is mechanically equivalent.}\\
A) An LCE kirigami is affixed on both ends normal to the director {\bf n} and opens up when heated from $25\C$ to $80\C$  the same as an elastomer kirigami under uniaxial extension (B). C) Experimental setup using a tensile apparatus with  LCE sheets immersed in water at temperature $T$ to measure $F(\delta)$ [see main text]. D) Experimental measurements $F(\delta)$ for LCE sheets with a single cut in the following situations: 
 uniaxial extension at fixed temperature $T=25\C$ (empty symbols) or fixed extension and increasing temperature (filled symbols) for cut lengths $l_0 =$ $10$\,mm (diamond), $20$\,mm (hexagon), and $30$\,mm (triangle). The black circle represents the regions where the cut openings are compared in E), with the LCE sheet at $25\C$ is under uniaxial force $F = 1.5 N$ (top), or with applied force $F = 1.3 N$ to maintain the rest length fixed while the temperature $T$ of the bath is changed from $25\C$ to $35\C$ (bottom). The agreement between the experimental protocols and cut geometry
indicates that the mechanics
 determines the cut opening, irrespective of the pathway (pulling or heating) that drives it.
 %For all panels,  {\bf n} indicates the orientation of the director of the LCE sheets at $25\C$. 
 Scale bars are 1cm.
 }
\label{fig:figure3}
\end{figure}

To investigate this analogy quantitatively, we study a single cut in an LCE sheet, under uniaxial load (at fixed temperature) 
or during temperature change (with fixed length) [Fig.~\ref{fig:figure3}D-E].  We perform experiments using a tensile apparatus with the LCE sheet fully immersed in water to control the sample temperature [Fig.~\ref{fig:figure3}C]. 
The cut opens under uniaxial stretch, or upon temperature increase from $25\C$ to $40\C$, beyond which it tears [SI Sec.~S3.5].
 %compromise between isotropic behavior and isotropic, experiments are performed at $XX \C$, above the order-disorder transition of the LCE [SI]. 
Experiments are performed as following: for uniaxial stretch,  we set the tensile force to zero when the cut is closed ($\delta=0\,$mm) and measure force $F$ vs opening $\delta$ of the cut during stretch, for increasing cut lengths $l_0$ [Fig.~\ref{fig:figure3}D, empty symbols]; for temperature-controlled opening, we set zero force with the closed cut ($\delta=0\,$mm) {\it at room temperature} and we measure the tensile force required to maintain this initial distance fixed as the temperature of the bath is increased, causing the LCE sheet to contract and the cut to open [Fig.~\ref{fig:figure3}D, filled symbols, Figs.~S6,~S7]. The heating of an LCE kirigami is a complex process,  
which includes changes both in the rest configuration of the material sheet and in the material properties themselves.
 Remarkably, the measurements of cut opening $\delta$ as a function of the force $F$ under mechanical extension %or
 or under thermal contraction agree [Fig.~\ref{fig:figure3}D], as well as the morphology of the cut openings [Fig. 3E].  
 We theorize from those observations that the mechanical response from the heating of a tethered LCE sheet can be simulated by an isotropic shell \cite{Ciarlet05, DolfinX23} under mechanical extension to the initial length, which we validate by predicting quantitatively the tensile measurements obtained experimentally at $40\C$ and $80\C$ [Fig.~S8]. 
 %, an observed experimentally at $80\C$ [Fig.~\ref{SI-fig:si2}]. 
% Simulations quantitatively agree with the the force-strain obtained experimentally at $80\C$, where LCE is fully isotropic and satisfactory capture results at $40\C$, where LCE present elastic \TODO{[Fig.~\ref{SI-fig:si2}]} and optical  [Fig.~\ref{SI-fig:siPOLA}] anisotropy. 
This proves that the opening of the cut that arises from the contraction of the LCE sheet is mechanically and geometrically equivalent to the mechanical extension of the isotropic sheet with the same Young modulus. This allows us to resort to simplified mechanical models and establishes a pathway-independence principle. Geometry determines the mechanical state irrespective of whether tension arises mechanically or thermally. It is the basis for our proposed approach, turning the molecular phase transition into soft robotics applications.
  
%We now turn to cut patterns that program structures deployable in 3D and whose Gaussian curvature using temperature patterns and design grippers actuators, whose radial pressure is controlled.

\subsubsection*{SuperSoft Grippers}
%We first engineer supersoft grippers with remote actuation. 
%We first control the effective elasticity of an LCE sheet by altering cut patterns at $80\C$, where the LCE sheet is mechanically isotropic [Fig. Sxx]. 
Similar to conventional kirigami, the addition of cuts decreases the stiffness of the LCE kirigami [Fig.~S9A]. 
We exploit this effect 
to design grippers with tunable grip strength and remote actuation [Fig.~\ref{fig:figure4}A]. LCE kirigami are rolled into a cylinder shape, with their ends attached to one another and placed upon a solid object whose circumference is below the length of the LCE at $T=25\C$. Upon temperature increase, the LCE sheet contracts, ends pulling onto one another: the cuts open, and the material presses inwards. This radial pressure $P_r$ is computed from pressure vessel theory \cite{pressure_vessels}: 
$P_{r} = \frac{1}{{2\pi R w}}T_0$, where $T_0$ is the tension exerted upon itself by the LCE kirigami, $R$ is the radius of the wrapped object and $w$ is the width of the LCE kirigami.  First, the tension $T_0$ is hereby bounded by the tensile force $F$ [Fig.~S9A] and controlled by the contraction of the LCE under heating. 
%Following our results [Fig.2A], LCE kirigami constitute more gentle grippers than uncut LCE sheets. 
Second, the cuts patterns decouple the tension $T_0$ and width $w$, enabling \MS{constrictive pressure down} to a few kPa [Fig.~S9C]. 
This makes it possible to engineer remotely actuated supersoft grippers, capable of grabbing a fragile object such as a thin shell made of aluminum foil (thickness $\sim16$\,\micro m and buckling threshold $\sim10$\,kPa, [Fig.~S9B-C, movie S2]. The LCE kirigami is placed upon the shell at room temperature, bracing it when immersed in hot water. The shell and gripper are subsequently displaced and during cooling, the LCE kirigami expands again and the shell is released. Inspection of the aluminum shell after this manipulation shows no signs of damage [Fig.~\ref{fig:figure4}A], confirming that the LCE kirigami gripper did not exceed the buckling pressure. In contrast, uncut LCE sheets or single threads result in crumpling of the shell [movie S2]. This makes LCE kirigami an attractive and programmable choice for supersoft grippers, combining remote actuation and manipulation of fragile objects, both challenges in robotics \cite{gripper_ultragentle, kirigrip, kiricurvature,doi:10.1073/pnas.1003250107}.
 %Practically, this makes LCE kirigami an attractive choice for soft robotics, combining remote actuation and manipulation of fragile objects. 
%LCE kirigramiss with fewer cuts as in [Fig. 2B], uncut LCE sheets or LCE threads all crumple the aluminium shell after contraction under the structure [Fig.2E. H, Fig. Sxx].  
%crumpled by impact of a simple pea  or manipulation by a uncut LCE sheet [Fig. Sxx].

%\subsection*{Macroscopically-controlled Geometry}
\subsubsection*{Programmable 3D structures}
\begin{figure}[tbhp]
\centering
\includegraphics[width=0.9\linewidth]{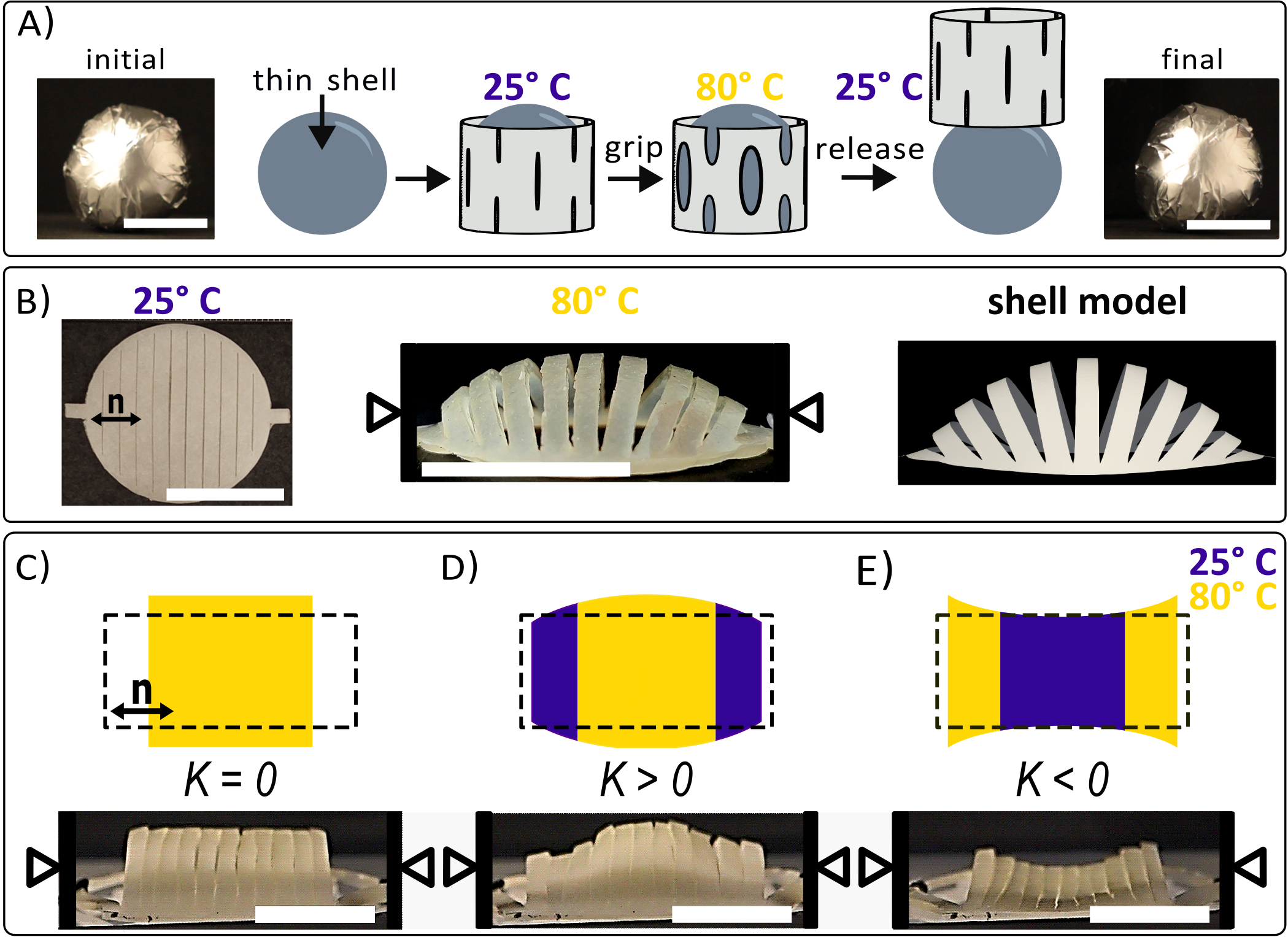}
\caption{
{\bf  Molecular phase transition of LCE kirigami controls soft robotics functions.}\\
A) {\bf SuperSoft Gripper} An LCE kirigami is rolled and tethered to its own end to form a SuperSoft gripper capable of handling fragile objects, here an aluminum shell of thickness $\sim 16\mu m$ (initial image to the left). The LCE kirigami contracts when heated from $25\C$ to $80\C$, bracing the shell and gripping. The cut pattern is selected so that the radial pressure on the shell is below the buckling threshold ($\sim 10~\textrm{kPa}$) [see main text]. After cooling down to $T=25\C$, the LCE kirigami releases the grip and the shell is undamaged (final image, right). See also [Movie S2].
B) A 2D LCE kirigami with positive boundary curvature (left, top view) is tethered on both ends ($T=25\C$) (as indicated by the black triangle). It deploys in a 3D structure with Gaussian curvature $K>0$ (middle, side view) following temperature increase to $T=80\C$, i.e. "pulling by heating"  and predicted by the simulation with isotropic material (right). {\bf C-E) Programmable 3D structures from a single rectangular precursor using temperature patterns:} A 2D Rectangular LCE precursor (dashed lines) presents zero boundary curvature at $T=25\C$. The boundary curvature is controlled by temperature patterns (blue: $25\C$, yellow: $80\C$, top view). Tethered LCE kirigami deploy reversibly in 3D in structures with C) zero, D) positive, or E) negative boundary curvature (side view),  temperature field as schemed above and measured on [Fig. S10]. Scale bars are 2 cm.
} 

\label{fig:figure4}
\end{figure}

Finally, we show that the molecular phase transition coupled to cut patterns programs macroscopic 3D deformations of LCE kirigami. For this, we take inspiration from conventional kirigami sheets, where 2D precursors with different boundary curvatures deploy into 3D shapes under uniaxial tension \cite{curvaturekiris}, an effect originating from the classical Gauss--Bonnet theorem that links boundary curvature with the global Gaussian curvature  $K$ \cite{ddgGauss-Bonet}. %It enables the formation of 3D morphologies from curved boundaries rather than complex cut patterns. 
Similarly, LCE kirigami deploy in 3D under remote actuation by leveraging \MS{their unique "pulling by heating-up"} feature [Fig.~\ref{fig:figure4}B-E]. We obtain a 3D dome from a 2D precursor with positive boundary curvature [Fig.~\ref{fig:figure4}B] as predicted by direct extension-based simulations.
%with negative, zero and positive boundary curvature respectively, deploy under tension into 3D shapes with negative, zero and positive Gaussian respectively. Hereby, we demonstrate similar behavior under remote actuation by leveraging the opportunity that LCE kirigami present from "pulling by heating" [Fig.3A-B] as further described by our extension-based simulations [Fig.3C].
Next, as LCE kirigami possess both the contraction and orthogonal expansion required to change boundary curvature,  we engineer temperature patterns [Fig. S10] to control the curvature directly and program different structures from a {\it single}  precursor instead of a set of them, like for conventional kirigami. A rectangular LCE kirigami deploys reversibly into multiple 3D targets with zero [Fig.~\ref{fig:figure4}C], positive [Fig.~\ref{fig:figure4}D], and negative [Fig.~\ref{fig:figure4}E] Gaussian curvature $K$ by changing the temperature patterns [movie S3]. 
%\JP{
%The interplay of the cuts with the molecular order of LCEs reversibly activates additional degrees of freedom.
%} 
%\MS{
This approach offers a versatile toolbox that complements previous realizations of programmable LCE architectures \cite{bilayer_LCE, LCE_lattice} and resolves a challenge in the field of metamaterials by programming a single kirigami sheet into multiple 3D structures  \cite{kiricurvature, reconfig_kirigami}.
%} 
% making them overcome the need for specific 2D templates as for elastomer kirigami \cite{kiricurvature} or spatial control of nematic director for conventional LCE robots. 
%\JP{
It further unlocks opportunities for inverse design and heating patterns towards complex targets and dynamics that remain to be investigated.
%}

%, that expands perpendicular to the director in regions that are heated [Fig.~\ref{fig:figure3}C], ([Fig.~Sxx for thermal imaging]). Following, a single LCE rectangular precursor deploy reversibly into multiple 3D targets programmed by  spatiotemporal temperature patterns [Fig. 3D]. The inverse and direct design of more complex targets and the required heating patterns remain to be investigated but are beyond the scope of this work. 
%\ML{a positive boundary curvature is obtained by heating up the central region and, conversely, negative boundary curvature can be obtained by heating up the ends [Fig.~\ref{fig:figure3}E], ([Fig.~Sxx for thermal imaging].}

%In addition, the molecular anisotropy of LCE sheets at room temperature leads to non-trivial viscoelastic behaviors and dynamical features, which we explore in the following.
%LCE sheets are anisotropic at room temperature a result of the molecular order of their liquid crystal mesogens, resulting in viscoelastic behavior. This inspires us to investigate the dynamic properties of LCE kirigami. %This anisotropy  results in viscoelastic behavior of LCEs before the transition, beyond which they recover isotropic behavior. 
%We demonstrate that cut patterns interact with the anisotropic properties of uniaxial LCEs sheet and provide a macroscopic lever to tap into the molecular properties of LCEs. We leverage those to unveil and rationalize dynamical properties of LCE kirigami, without counterpart with conventional kirigami.

\section*{Discussion}
%\JP{
We showed that patterns of cuts in LCE sheets amplified and controlled molecular anisotropy, enabling emergent mechanical behaviors that are inaccessible to conventional kirigami or LCEs alone. 
%} 
LCE kirigami embody multiscale metamaterials, coupling mesoscopic cut patterns and molecular properties to become more than the sum of their parts. The dynamics of relaxation after stretch reflect the interplay between the redirection of strain by cuts and the molecular order of LCEs.
%Macroscopic cuts in LCE sheet redirect and concentrate strain, tapping into the molecular anisotropy of the LCEs.
%\JP{
The mechanism requires anisotropy and strain-dependent relaxation, hence could generalize to other viscoelastic materials or responsive materials, such as ambidirectional LCE \cite{doi:10.1126/science.adq6434}, fibrous materials \cite{paper, wood}, or collagen-based tissues \cite{Fratzl2008}.  
%}
We validated a computational pipeline that establishes the recoil dynamics from static computation of mechanical strain in an elastic kirigami sheet. Our understanding of the interplay  between mechanics and dynamics, macroscopic and molecular properties, allows the programming of shapes of LCE kirigami by stretch and release.   
Cuts transform the molecular phase transition of LCEs into a powerful design instrument for soft robotics, making it possible to design supersoft grippers or multi-shape deployable 3D structures, opportunities beyond conventional kirigami or  LCE sheets only.
Altogether, our results show that cut patterns are an accessible, potent, and versatile tool to program mechanical and soft robotics functions. They couple mesoscopic and molecular features, bridging scales from the centimeter down to the nanometer.  We further envision this approach to benefit from  technological advances in director patterning at the microscale \cite{10.1126/sciadv.abc0034, leaplce}, the adaptation of LCEs by training \cite{newLCEmetamat}, and progress in direct or inverse design to program hierarchical and multiscale metamaterials made of anisotropic and responsive materials. LCEs also provide an opportunity to power and reconfigure metamaterials remotely, a step towards active metamaterials. Complex cut patterns, such as spirals, could lead to rotational motion under contraction, an elemental motif of odd elastic behavior and an uncharted route to power robots with non-reciprocal interactions \cite{10.1038/s41586-025-08646-3}.

%%%%%%%%%%%%%%%% ACKNOWLEDGEMENTS %%%%%%%%%%%%%%%
\paragraph*{Acknowledgments:}
We thank L. Bocquet for discussion and feedback. We thank C. Farr and C. Plamadeala for assistance in the kirigami fabrication, M. Lorion and M. Bhargava for the initial support with the material synthesis, and C. Hafner and C. Wojtan for feedback and discussion on the computational components.
This project has received funding from the European Research Council (ERC) under the
European Union’s Horizon Europe
research and innovation program (VULCAN, 101086998). Funded by the European Union. Views and opinions expressed are however those of the authors only and do not necessarily reflect those of the European Union or the European Research Council Executive Agency. Neither the European Union nor the granting authority can be held responsible for them. JP thanks the Machine Shop (MIBA) of ISTA's Shared Scientific Units (SSU) for their support.

\paragraph*{Author contributions:}
MS and JP conceived the project, with feedback from BB. MS and QM designed and performed the experiment. ML and MS developed and performed the simulations. MS and JP wrote the manuscript. All authors contributed to the SI. All authors reviewed and commented the manuscript.

\paragraph*{Competing interests:}
The authors declare no competing interests.

\subsection*{Supplementary materials}
Captions for Movies S1 to S3\\
Movies S1 to S3
Supplementary Text\\
Figures S1 to S25

\clearpage

% \newpage
% \begin{figure}[tbhp]
% \centering
% \includegraphics[width=\linewidth]{figure5.pdf}
% \caption{
% \JP{
% {\bf:  Programming the recoil direction of asymmetric LCE kirigami.}  
% (a) Pattern of cuts used for the experiments (top). Asymmetric cuts accumulate orthogonal strain $\epsilon_{YY}$ on one side of the kirigami under uniaxial extension as visible from FEM simulations (bottom). (b) For LCEs at $80\C$ or isotropic elastomers (top), the retraction time decreases with strain [triangles Fig. 4C], the side with higher $\epsilon_{YY}$ retracts faster leading to a bending of the LCE towards this side after extension. In contrast, the retraction time increases with strain for LCEs at room temperature [circles, Fig. 4C], the side with higher $\epsilon_{YY}$ retracts slower leading to a bending in the opposite direction than for the elastomer (bottom). (c) The  angle of retraction is controlled by the temperature of the LCE, reversing at 80$\C$ for isotropic LCEs. Transparency scale indicates time of retraction, from near transparent for the initial position to darker during the retraction. (d) Control of the retraction angle, as defined on the figure and dependence on the temperature.
% } 
% }
% \label{fig:figure5}
% \end{figure}

\clearpage

\bibliographystyle{unsrt-custom}
\bibliography{ref.bib}

\end{document}